\documentclass[notitlepage,preprint]{article}
\usepackage[bitstream-charter]{mathdesign} 
\usepackage[colorlinks=true,urlcolor=blue,linkcolor=blue, citecolor=blue]{hyperref}
\usepackage{graphicx}
\usepackage[a4paper,left=3.5cm, right=3.5cm, top=3cm, bottom=3cm]{geometry}
\usepackage{authblk}

\usepackage{amsthm}
\usepackage{amsmath}

\title{Implementing the De-thinning Method for High Energy Cosmic Rays Extensive Air Showers Simulations}

\author[1]{Alex Estupi\~nan\thanks{Corresponding author: \href{mailto://alex.estupinan@correo.uis.edu.co}{alex.estupinan@correo.uis.edu.co}}}
\author[2,1]{Hernan Asorey}
\author[1]{Luis A. N\'u\~nez}

\affil[1]{Escuela de F\'isica, Universidad Industrial de Santander,
Bucaramanga, Colombia}
\affil[2]{Laboratorio Detecci\'on de Part\'iculas y Radiaci\'on, Centro At\'omico Bariloche \& Instituto Balseiro, San Carlos and Bariloche, Argentina}

\date{\today}
\begin{document}
\maketitle

\begin{abstract}
To simulate the interaction of cosmic rays with the Earth atmosphere requires highly complex computational resources and several statistical techniques have been developed to simplify those calculations. It is common to implement the  thinning algorithms to reduce the number of secondary particles by assigning weights to representative particles in the evolution of the cascade. However, since this is a compression method with information loss, it is required to recover the original flux of secondary particles without introduce artificial biases. In this work we present the preliminary results of our version of the de-thinning algorithm for the reconstruction of thinned simulations of extensive air showers initiated by cosmic rays and photons in the energy range $10^{15} < E/\mathrm{eV} < 10^{17}$.
\end{abstract}

\section{Introduction}

The Earth atmosphere interacts continuously with a flux of particles of galactic and extragalatic origin. The interaction of one of these cosmic rays with an atomic element in the atmosphere produces a cascade of particles (the so called Extensive Air Shower, EAS), in which the number of particles could reach billions at the maximum of the development of the cascade.

To understand and simulate these processes requires highly complex computational resources. To reduce the number of particles that have to be followed at the highest energies, it is common the usage of the {thinning} algorithm (see for example \cite{Billoir2008}), a statistical method that reduce the number of secondary particles by assigning weights to representative particles in the evolution of the cascade. However, since this is a compression method with loss of information, it is required to recover the original flux of secondary particles without introduce artificial biasses. 

The so called {de-thinning} method \cite{StokesEtal2012} is one of the existent methods designed to deal with this information loss. Since the EAS is an stochastic process, in this work we will compare the distribution of particles at ground level for several non-thinned showers, with the corresponding reconstructed thinned (with two different levels of thinning, $\epsilon_{th}=10^{-7}$ and $\epsilon_{th}=10^{-8}$) and de-thinned showers obtained from exactly the same set of initial parameters. All the simulations were made by using CORSIKA air shower simulation program \cite{HeckEtal1998} and several analysis routines developed within the LAGO (Latin American Giant Observatory) project simulation chain\cite{Asorey2013b,Asorey2014a}.

\section{Thinning Method}

The implementation of this algorithm is applied during the shower simulation over the secondary particles when this condition is fulfilled: 
\begin{equation} 
E_{0}\epsilon_{th} > \sum_{j=1}^{n} {E_j}, 
\end{equation}
where $E_j$ is the energy of the secondary particle, $E_{0}$ is the energy of the primary particle and $\epsilon_{th} = E_j/E_0$ is defined as the level of thinning.

In this case, only one secondary particle $i$ survives probability:
\begin{equation} 
P_i = E_i / \sum_{j=1}^{n} {E_j}.
\end{equation}
Otherwise, if the total sum of the energy of the $n$ secondary particles is greater than the thinning energy threshold, i.e.:
\begin{equation} 
E_{0}\epsilon_{th} < \sum_{j=1}^{n} {E_j},            
\end{equation}
then the secondary particle with energy below the thinning threshold will survive with a probability:
\begin{equation} 
P_i = E_i / E_{0}\epsilon_{th}.
\end{equation}
In both cases, the particles that survive have their weight multiplied by a factor of $w_i=P_i^{-1}$. 

\section{De-thinning Method}

The main questions to answer for any reconstruction method of thinned showers are:

\begin{itemize}
\item How do you determine the accuracy of the survival sample of secondary particles?
\item How do you use the thinned sample to completely rebuild the original shower avoiding the introduction of artificial biases?
\item What is the maximum value of $\epsilon_{th}$, for which the sample is a good representative of the original shower for a particular type of reconstruction method?
\end{itemize}

Several techniques has been developed to asses these questions. Our implementation of the de-thinning algorithm is based in the original development of Stokes et al. \cite{StokesEtal2012}, and consist in the successive application of the following steps over the thinned sample of secondary particles: 
\begin{enumerate}
\item choose a vertex point on the trajectory of the weighted particle in the way given in the next paragraph;
\item centred in the weighted particle path, chose a random direction of propagation for the new particles (also called ``daughter'' particles) by using a 2D gaussian distribution with zero mean and a fixed standard deviation; 
\item project the daughter particle to ground level, and calculate the travel distance in units of atmospheric depth;.
\item determine the energy of the inserted particle using a gaussian random distribution with mean $E_i$ and $\sigma=0.1 E_i$, where $E_i$ correspond to the energy of the weighted particle;
\item determine the probability of atmospheric absorption of the inserted particle by using a fixed atmospheric interactions length, and decide if it will reach the ground;
\item depending on the daughter particle trajectory and its starting point, calculate the time of flight of the inserted particle;
\item Repeat steps two to six, $(w-1)$ times to build a complete set of secondary unweighted particles.
\end{enumerate}

To assure temporal consistency of the particles reaching ground level, the distance between the vertex where the sub-shower of the de-thinned particles begins and the ground should be less than $D_{max}$, which is given by: 
\begin{equation}
D_{max}=\frac{c^{2}(t_{i}-t_{0})^{2}- \mid \vec{x_{i}}-\vec{x_{0}}\mid^{2}}{2(c(t_{i}-t_{0})-(\vec{x_{i}}-\vec{x_{0}})\cdot\hat{p_{i}})}
\end{equation}
where $c$ is the speed of light. Any shorter separation will generate de-thinned particles temporally consistent with the development of the EAS. 

We have implemented the above de-thinning algorithms, and introduce some improvements in the overall algorithm, such as:
\begin{itemize}
	\item the standard deviation used to determine the daughter trajectory depended on the type (electromagnetic, muon or hadron) and the energy $E_i$ of the weighted particle;
    \item location dependent atmospheric model to determine the atmospheric depth as a function of the altitude and trajectory of the weighted particle;
	\item the atmospheric interaction length used to determine the probability of reaching ground level also depends on the type of secondary particle;
\end{itemize} 
Our de-thinning code was implemented in python 2.7\footnote{\href{https://www.python.org/}{www.python.org}} and will be licensed under GPLv3\footnote{\href{https://www.gnu.org/copyleft/gpl.html}{www.gnu.org/copyleft/gpl.html}}.

\section{Results}

\begin{figure}
\begin{center}
\includegraphics[scale=.18]{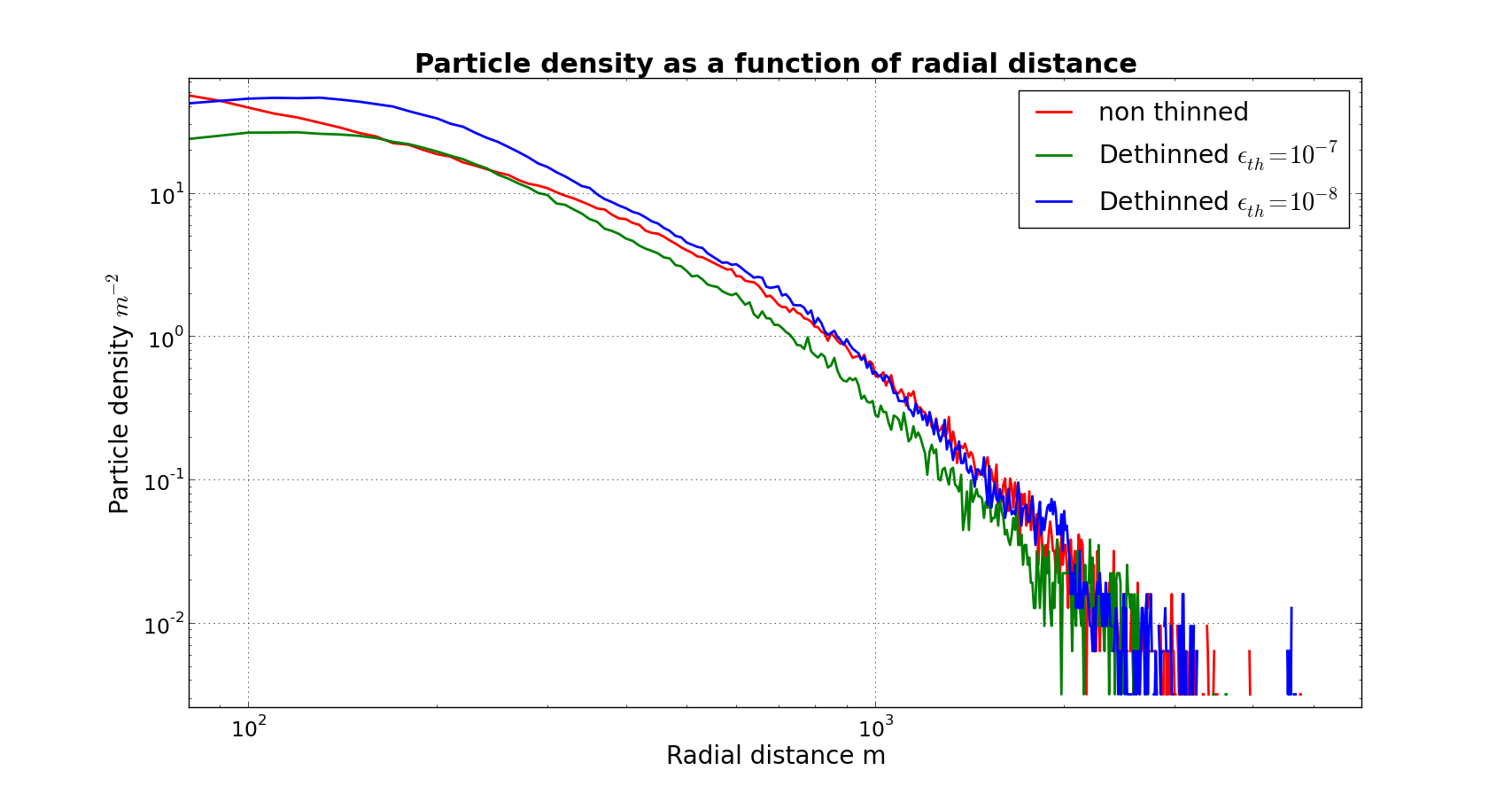}
\caption{Particle density as a function of the core distance for a vertical proton of energy $5\times 10^{6}$\,GeV. We use the non-thinned shower (red line) as the reference, and compare it with the reconstructed density profile obtained after the application of our de-thinning method over the same showers but with thinning levels of $\epsilon_{th} = 10^{-7}$ (dark green line) and $10^{-8}$ (blue line). We optimize our method to get good results on intermediate distances to the core, where typical detectors used in many astroparticles observatories are triggered but not saturated.}
\label{density_part}
\end{center}
\end{figure}

\begin{figure}
\begin{center}
\includegraphics[scale=.14]{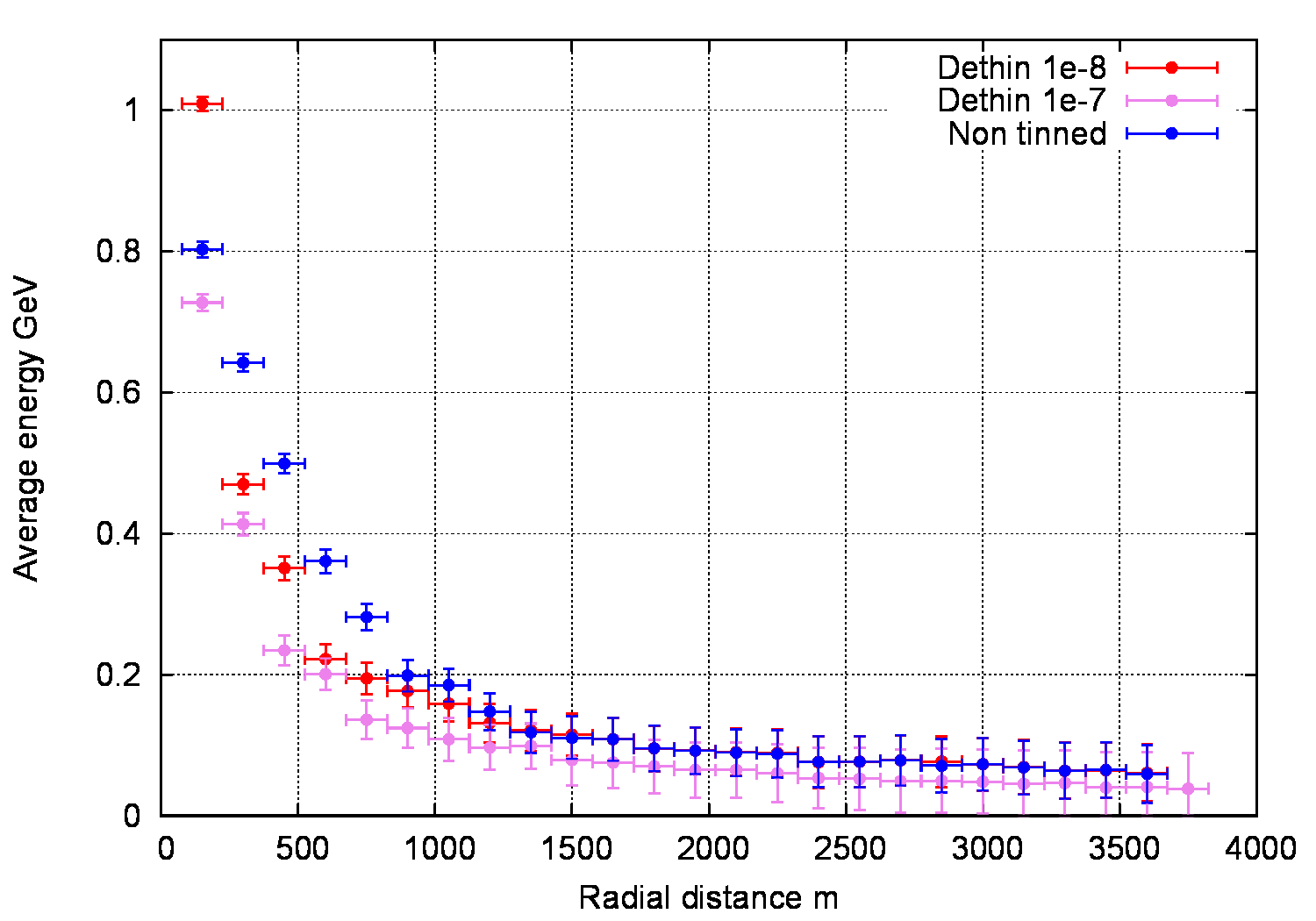}
\caption{Average energy of secondaries as a function of the radial distance for a shower initiated by a vertical proton with energy $E_0=5 \times 10^{6}$\,GeV without thinning (blue circles), and de-thinned with original thinned levels of $\epsilon_{th} = 10^{-7}$ (magenta thin circles) and $\epsilon_{th} = 10^{-8}$ (blue circles).}
\label{avg-energy}
\end{center}
\end{figure}

Our main results for this first approach to this problem are displayed in figures \ref{density_part} and \ref{avg-energy}, where it is possible to compare the secondary particle density and the average energy as a function of the distance to the shower core. As explained above, we simulate showers at fixed energies and arrival directions for different primaries and different values of thinning levels: $\epsilon_{th} = 0$ (non-thinned, used as reference) and $\epsilon_{th} = 10^{-7}$ and $\epsilon_{th} = 10^{-8}$. The set of parameters used for each simulation, including the CORSIKA random generators seeds, were exactly the same to allow proper comparisons between the de-thinned and the non-thinned showers.

With this implementation, we obtained very good results with a significant improvement in the usage of computing resources, which is specially useful for simulate large numbers of high energy showers in remote simulation clusters or GRID-CORSIKA implementations, as the cloud storage and network transfers of the large outputs obtained without thinning could be extremely difficult. The outcomes of the computing times and the storage for the different configurations tested are presented in the table \ref{tabla1_1}. It is clear a significant reduction in the computing times and output file sizes.

\begin{table}[!h]
\centering
\begin{tabular}{| c | c | c | c | c |}
	\hline 
   $\mathbf{\epsilon_{th}}$ & \textbf{None} & $\mathbf{10^{-8}}$ & $\mathbf{10^{-7}}$ & $\mathbf{10^{-6}}$ \\ 
	\hline
   \textbf{Comp. time (h) }    & 48.92      & 7.42        &    5.69 &  1.12  \\ 
   \textbf{Storage (MiB)}  & 895  & 750     & 645 & 431 \\  
	\hline
	\multicolumn{5}{c} \textbf{Reconstructed showers used our de-thinning implementation}  \\ 
	\hline
    \textbf{Comp. time (h)}  & --      & 7.74       &    5.81 &  1.19  \\
    \textbf{Final storage size (bz2, MiB)} & -- & 753 & 725 & 731 \\
	\hline 
    \end{tabular}
	\caption{Simulation time and final secondary particles file sizes for different thinning levels ($\epsilon_{th}$) for a shower initiated by proton of $E_{0}=10^{8}$\,GeV.}
\label{tabla1_1}
\end{table}

\section{Conclusions and Acknowledgements}

In this work we show the first results of a python-based implementation of the
so called de-thinning algorithm, which is applied for the reconstruction of
thinned CORSIKA simulations of the interaction of primary cosmic rays with the
atmosphere. After comparing the results of the application of this algorithm
over thinned showers with the corresponding non-thinned equivalent ones, we
were able to tune different parameters of the reconstruction method, such as
the energy dependent angular aperture of the cone for the injected new
particles and specific atmospheric absorption coefficients for each type of
secondary particle. When totally implemented, this code will be publicly
released under the license GPLv3.

The authors of this work thank the support of COLCIENCIAS ``Semillero de
Investigaci\'on'' grant 617/2014 and one of us (LAN) to CDCHT-ULA project
C-1598-08-05-A.

\bibliographystyle{elsarticle-num}
\bibliography{biblio}
\end{document}